\documentstyle[11pt]{article}
\begin{document}

\title{Holographic Principle bounds on Primordial Black Hole abundances}

\author{P.S.Cust\'odio and J.E.Horvath\\
\it Instituto de Astronomia, Geofisica e Ciencias Atmosfericas\\ Rua do
Mat\~ao 1226, 05508-900 S\~ao Paulo SP, Brazil\\ Email:
foton@astro.iag.usp.br}
\maketitle

\maketitle

\abstract{ The generalized Second Law of thermodynamics and the
Holographic Principle are combined to obtain the maximum mass of
black holes formed inside a static spherical box of size $R$
filled with radiation at initial temperature $T_{i}$. The final
temperature after the formation of black holes is evaluated, and
we show that a critical threshold exists for the radiation to be
fully consumed by the process. We next argue that if some form of
Holographic Principle holds, upper bounds to the mass density of
PBHs formed in the early universe may be obtained. The limits are
worked out for inflationary and non-inflationary cosmological
models. This method is independent of the known limits based on
the background fluxes (from cosmic rays, radiation and other forms
of energy) and applies to potentially important epochs of PBH
formation, resulting in quite strong constraints to
$\Omega_{pbh}$.}

\maketitle

\section{Introduction}

\bigskip

It is a known fact that small primordial black holes (PBHs) are
not abundant today because these objects evaporate quickly due to
Hawking radiation [1]. Rather strong limits to their relative
abundances have been obtained , as discussed in several
contributions in Refs. [1-4]. An appraisal of these limits has
been recently given in Ref.[2], in which the constraints for
several initial PBH masses and formation scenarios have been
addressed.

Generally speaking, all these methods are based on comparisons of
the integrated contribution of Hawking radiation with some
background flux and related arguments, such as deuterium abundance
and $^4$He spallation limit. Although the figures depend on the
considered interval of masses, it is fairly general to say that
present limits suggest $\Omega_{pbh}<{10}^{-6}$ for PBHs with
masses $\leq$ the Hawking mass $M_{haw}$ (defined as the
mass-scale which is evaporating precisely today). We will here
explore another complementary method to limit PBHs abundances
based on ideas of entropy bounds.

It is well known that black holes possess some peculiar
properties. For instance, they are the most entropic objects for a
given mass and radius found in nature. Expressed in units of the
Boltzmann constant $k_{B}$, a black hole has a huge entropy given
by the horizon area as

\begin{equation}
S_{bh}(M)={A\over{4}}\sim{10}^{77}{(M/M_{\odot})}^{2} \, .
\end{equation}

It is important to note that the total observed entropy of the universe is
just about 10 orders of magnitude greater than a $M \sim M_{\odot}$
black hole, some $S_{0}\sim
{10}^{87}$, contained mainly in the radiation originated in the
Big Bang.

The origin and interpretation of this huge entropy is a
complicated issue and its complete understanding will establish an
important bridge between quantum mechanics, general relativity and
thermodynamics. The main problem lies in a statistical
interpretation of this entropy. In statistical mechanics, the
entropy is associated  to the counting of quantum microstates of
the system as given by $S \propto \log{\Omega}$. But which are the
quantum microstates of a black hole and how can we count them?
These interesting interpretation problems were analyzed in
Bekenstein [6], Jacobson [7] and several recent works which
address the nature of black hole entropy and attempt a more
comprehensive understanding (see [8] for recent reviews).

It is the purpose of this work to obtain conditions for PBH
formation using entropy concepts. Next section reviews some
general concepts such as the Holographic Principle and the
generalization of the second law of thermodynamics, to be needed
later. Section 3 discusses the case of a closed box in which PBH
formation is induced by external perturbations in order to show
how these bounds work. Section 4 presents a brief discussion on
the preferred final state. In section 5 non-inflationary
cosmological scenarios are addressed and limits to the PBHs
abundance derived. Section 6 discusses the same problem in
presence of inflation and bounds on the physical temperature for
PBHs to form. Conclusions are the subject of Section 7.

\bigskip

\section{Entropy bounds and the Generalized
Second Law of thermodynamics}

\bigskip

The Holographic Principle states that for a given volume $V$, the
state of maximal entropy is proportional to the area $A$ bounded by the
volume $V$. The microscopic entropy $S$ associated with the
volume $V$ is bounded by the Bekenstein-Hawking entropy

\begin{equation}
S_{BH} \leq{A\over{4}}
\end{equation}

in which the Planck length $L_{p}$ has been set to unity. Verlinde
[9] observed that this bound has to be modified in the
cosmological setting with arbitrary dimensions. That work
illuminated some interesting connections between entropy and
dynamics; and showed that different forms of cosmological entropy
bounds may be closely related; namely the Bekenstein-Hawking form
(eq.2), the Bekenstein-Verlinde bound
$S_{BV}={2\pi\over{n}}ER$(with $n$ the number of spatial
dimensions) possibly interpreted as a consequence of the quantum
Principle of Uncertainty [10] ; and the Hubble bound
$S_{H}=(n-1){HV\over{4G_{n}}}$ (with $G_{n}$ the generalized
Newton constant). At the point in which the Hubble parameter times
the radius $R$ satisfy $HR =1$ the Friedmann-Robertson-Walker
equations yield a value of the entropy akin to the Cardy-Verlinde
formula [9]. These are strong hints (but not rigorous proof) that
the dynamics of the cosmological model has a built-in knowledge of
the entropy within its limits.

A second important concept to be used below is the generalized
Second Law of thermodynamics, first formulated by Bekenstein in
70s [11]. In a series of gedanken experiments, Bekenstein noted
that there were serious problems with the matter/radiation entropy
as it was absorbed onto black holes (causing a mass growth of the
black hole that absorbs that matter/radiation). However, a series
of powerful no-hair theorems granted that black holes are
described by very few macroscopic parameters (mass, angular
momentum and charge), and after absorption of matter, it is not
possible to retrieve the initial conditions and describe the
quantum numbers of the absorbed matter incident onto this object.
In other words, Bekenstein noted that as black holes absorb
matter, the entropy of the universe seems to decrease, since the
matter entropy vanished behind the horizon. This was quite
problematic, since this {\it Geroch} process seems to go against
the second law of thermodynamics $\Delta{S_{matter}}>0$.

In order to recover the second law of thermodynamics, Bekenstein
[11] conjectured a Generalized Second Law of Thermodynamics (or
GSL), suitable to be applied to black holes and the
matter/radiation outside of these objects. According to this
conjecture the total entropy of the universe plus $N$ black holes
is given by the sum of the matter/radiation entropy plus black
hole entropies (which are proportional to their horizon areas).
Then, the GSL takes the form

\begin{equation}
S_{total}=S_{m-r}+{1\over{4}}\sum_{i}^{N}A_{i}
\end{equation}

As long
as we deal with classical process involving black holes and
matter, the total variation of entropy must be positive

\begin{equation}
\Delta{S_{total}}>0
\end{equation}

The GSL can be used analogously to the well-known use of the
second law for ordinary systems. In the section 3 below, we shall
use eq.(2) together the GSL for a given box with a well-defined
energy and radius to show how to obtain upper limits to the black
hole masses formed inside the box. In sections 5 and 6 we use the
Holographic Bound, $S_{total}<{A\over{4{L_{pl}^{2}}}}$,
conjectured to be valid for weakly or strongly gravitating
systems, to repeat the exercise and show the limits for the case
of realistic cosmologies.

\bigskip

\section{Black hole masses allowed in a closed box}

\bigskip

Before considering the case of realistic cosmologies let us begin
with a simpler system, namely a finite closed box with thermal
radiation in which black hole formation is induced. In the first
gedanken experiment this closed ideal box is filled with thermal
radiation at initial temperature $T_{i}$ and no black holes at
all. Then, we allow that this system to pass through a finite
region with strong metric perturbations. The final state of the
experiment is a box containing thermal radiation at some final
temperature and black holes with a particular Initial Mass
Function (hereafter IMF), the details of the latter depending on
the perturbed region. This is known as a Susskind process and it
is discussed in [12].

We are not interested here in the intermediate stages, because it
is enough to evaluate the initial and final entropies, taking into
account the conserved energy, and compare it with the Holographic
Bound. Then, we will assume that exists some particular metric
perturbations (acting over a finite time) in this box in order to
initiate the collapse of some fraction of initial energy in black
holes. Let us suppose that this perturbation gives rise to $N$
black holes with same initial masses $M_{i}$. By further imposing
that $V_{box}>>\sum_{i}^{N}{r_{g}}^{3}$ (where $r_{g}$ is the
gravitational radius of the black hole), we shall avoid complex
interactions between them.

Since we are not modelling the perturbations themselves, we need
this condition to assure that the radius and total energy of our
boxes are well defined. In practice, we will not explicitly use
these constraints, since that we only need to require the
background to be asymptotically flat. The complete requirements
for PBH formation are then two, enough primordial perturbations
and enough entropy for the final state (black holes plus
radiation). The condition of having enough perturbations can fix
only the functional form of the initial mass function of the PBHs
(IMF). The latter condition (enough entropy) completely defines
the IMF, determining the lower and upper masses cut-off in the IMF
(see below).

The Holographic Principle says that a finite system has a maximal
total entropy specified by its radius and energy, and the
Bekenstein-Verlinde limit adopts the numerical value

\begin{equation}
S_{BV} = {2\pi\over{n}}ER \sim {2\times{10}^{38}(M/g)(R/cm)}
\end{equation}

where  $n=D-1$ is the spatial dimension (hereafter set to $n=3$),
$M$ stands for the total mass-energy enclosed in this box and $R$
is its physical radius. The spherical box with radius $R$,
containing thermal radiation at temperature $T$ has an entropy
content given by

\begin{equation}
S_{rad}(T)=g_{*}(T){(TR)}^{3}
\end{equation}

and its numerical value is $\sim \, 2 \times {10}^{87}$ for
$T=T_{0}\sim{3K}$ and $R=R_{0}\sim{10}^{28}{h_{0}}^{-1} cm$
(the values chosen for the box to mimic the present universe).

We can write the entropy of the box in its final state, with $N$ black holes
plus leftover radiation as

\begin{equation}
S_{total}=NS_{bh}(M)+S_{rad}(T) < S_{BV}
\end{equation}

The total entropy variation is
the sum of the entropy variation of the total content of the box and the
entropy of the perturbations responsible for the black holes formation.
The irreversibility of the total process is expressed as

\begin{equation}
\Delta{S_{total}}=\Delta{S_{box}}+\Delta{S_{pert}}>0
\end{equation}

To calculate the complete outcome of the process we only need to know
the entropy and energy in the box,
since that the extensive parameters of the box are well-defined.

Provided $S_{box} < S_{BV}$ holds for all times, the formation of
the black holes is possible if the difference
$[S_{BV}-S_{rad}(T)]$ is positive, because black hole entropies
are positive. Then, using the Bekenstein-Verlinde Holographic
Bound $S_{BV} ={2\pi\over{3}}(E_{total}R)$ we obtain

\begin{equation}
{2\pi\over{3}}{ER}\geq{S_{rad}}(T)
\end{equation}

Considering $E_{total}=\varrho_{total}R^{3}$, and inserting above,
we obtain a constraint on the total density for black hole
formation to occur

\begin{equation}
\varrho_{total} \geq {3\over{2\pi}}{S_{rad}(T)\over{{R}^{4}}}
\end{equation}

(as we shall show in the next sections, an analogous relation holds for a
closed FRW universe when we consider a generalization of the
Holographic Principle to this cosmological situation)

Let us show how the formation of PBHs is characterized by the entropy and energy by
means of a simple explicit calculation.
The initial state of the box will be defined by

\begin{equation}
E_{i}=\varrho_{rad}(T_{i})V_{box}
\end{equation}

and

\begin{equation}
S_{i}(T)=g_{*}{T_{i}}^{3}V_{box}
\end{equation}

and the final state described by

\begin{equation}
E_{f}(M_{bh},T_{f})=NM_{bh}+\varrho_{rad}(T_{f})V_{box}
\end{equation}

and

\begin{equation}
S_{f}(M_{bh},T_{f})=g_{*}T_{f}^{3}V_{box}+f_{0}N{M_{bh}}^{2}
\end{equation}

with $f_{0}=2.5\times{10}^{10}$, the masses are measured in grams
and the temperatures in Kelvin degrees.
Using the conservation of energy, the radiation temperature
is found to fall as

\begin{equation}
T_{f}=T_{i} \times \xi(N,M)
\end{equation}

where we have defined the function
$\xi(N,M)={\biggl[{1-{c^{2}NM\over{E_{i}}}}\biggr]}^{1/4}$.
Since the total energy content in the box is
conserved, $E(t_{f})=E_{i}$, the eqs.(7),(13) and (14) can be combined
into a single expression of the form

\begin{equation}
B{M}^{2}+g_{*}V_{box}{T_{i}}^{3}\xi^{3}(N,M) \leq
{C{(E_{i}/c^{2})}{V_{box}}^{1/3}}
\end{equation}

with $B \sim {2.5\times{10}^{10}N}$ , $C \sim {2\times{10}^{38}}$
and $E_{i}=\varrho_{0}{T_{i}}^{4}V_{box}$ (with
$\varrho_{0}\sim{8\times{10}^{-36}g{cm}^{-3}}$).

We seek the maximum values of the formed black hole masses satisfying
the Holographic Principle. First, we note that exists an apparent
maximum for the black hole mass $M_{max}=(E_{i}/c^{2}N)$. In this
case, the final temperature of the radiation is zero. Then, the
final state consists of $N$ black holes with this maximum mass,
and an extreme cooling of the box happens, $T_{f}=0$.

However, for high temperatures, $T_{i} > T_{cr}$, (holding $N$,
$V_{box}$ and $M$ fixed) there will be an new local maximum for the
black hole masses, which clearly satisfies $M_{max}< E_{i}/{c}^{2}N$.
In order to obtain this new maximum, we impose that
$S_{f}(M,T_{f}) < S_{BV}$ and recall that
$E_{f}(M,T_{f})= NM_{bh} +\varrho_{rad}(T_{f})V_{box}$.
The entropy relation becomes

\begin{equation}
f(N,M) \leq {g(T_{f}(M),V_{box})}
\end{equation}

where $f(N,M)=(B{M}^{2} - CNM{V_{box}}^{1/3})$ and
$g(T_{f},V_{box})=(1.6 \times
10^{3}{T_{f}}^{4}{V_{box}}^{4/3}-g_{*}V_{box}{T_{f}}^{3})$.

On the curve $f(N,M_{*})= g(M_{*},V_{box})$ we find the
maximum value of the mass $M_{*}$ allowed by the HB. This
maximum will be {\it smaller} than the apparent value $E_{i}/{c}^{2}N$ if the initial
temperature of the box is larger than a critical value $T_{cr}$,
to be evaluated below. The situation is displayed in Figure 1.


When $T_{f} = T_{i}$ (i.e. in the limit of $M \rightarrow 0$ PBHs), the function
$g(T_{f},V_{box})$ is positive only for $T_{i} > {10}^{-3} {V_{box}}^{-1/3} K $. As
$M \rightarrow {M_{max}}$, the function $g \rightarrow {0}$, since
the final temperature $T_{f}$ also approaches zero.

We further note that $f(N,M)$ is positive for large black holes
and changes signs at smaller values. The global minimum of $f$ is
given by
$\biggl({\partial{f}\over{\partial{M}}}\biggr)_{M=M_{1}}=0$ ; and
leads to $M_{1}={CN \over{2B}}{V_{box}}^{1/3} \sim
{5\times{10}^{27}{V_{box}}^{1/3} g}$. The function $f$ adopts at
its minimum the value
$f(N,M=M_{1})=-{C^{2}N^{2}\over{4B}}{V_{box}}^{2/3}$ and
$f(N,M_{2})=0$ occurs for $M_{2}=2{M_{1}}$.

An inspection to Fig. 1 shows that the critical temperature (above which the formation of
black holes can not exhaust the radiation in the box)
must satisfy $2M_{1} < {E_{i}\over{c^{2}N}}$. Solving this
inequality we obtain for the critical temperature $T_{cr}(M,N,V_{box})$

\begin{equation}
T_{cr}(M,N,V_{box})={\biggl({2M_{1}c^{2}N\over{\varrho_{0}V_{box}}}\biggr)}^{1/4}
\sim{5.9 \times {10}^{15} {N}^{1/4}{(V_{box})}^{-1/6} K}
\end{equation}

These calculations lead us to conjecture that if a large number of
very large black holes were formed in the primordial universe,
they must have exhausted some fraction of the initial temperature
of the cosmic radiation due to the energy conservation. This is
simple to analyze in a closed finite box, and a similar reasoning
for the early universe can be carried over, albeit with several
mathematical complications. If we insist that
$\varrho_{pbh}<<\varrho_{rad}$ then the net effect on the
temperature must have been small, although such a sudden cooling
might have been sizable, for example, in the hypothesis of large
PBH formation as needed for seeds of the supermassive black holes
in the center of galaxies. The bottom line here is the belief that
the early universe was quite similar (thermodynamically speaking)
to a closed box filled initially with radiation, and we may choose
the particle horizon $R_{hp}(t)$ as a fiducial value for the
effective box that represents it. Therefore, these ideas developed
for boxes can be applied to the realistic universe with few
modifications in the reasoning (see sections 5 and 6).

\bigskip

\section{$N$ small black holes vs. one large black hole}

\bigskip

Before jumping to the analysis of cosmological models we can rise
the following question when considering perturbations that induce
black hole formation in the box. Which situation produces more
entropy? (and hence is more likely to occur), $N$ small black
holes (with identical masses, say $M_{*}$) or one very big black
hole (but also satisfying $R_{g}<< V_{box}$)? To address this
question let us prepare two boxes with the same initial
temperature $T_{i}$ and same volume. For a meaningful comparison
of the entropy variation, we must impose the same {\it final}
temperature. The variation of entropy inside the box for the first
case (one black hole formed) is

\begin{equation}
\Delta{S_{1}}=f_{0}{M}^{2}-{V_{box}\over{T_{i}}}
\biggl[{\varrho_{rad}(T_{i})\biggl(1-{1\over{\xi(1,M)}}\biggr)+{M
\over{V_{box}\xi(1,M)}}}\biggr]
\end{equation}

In the second case, the variation of entropy in the box is given by

\begin{equation}
\Delta{S_{2}}=f_{0}N{M_{*}}^{2}-{V_{box}\over{T_{i}}}
\biggl[{\varrho_{rad}(T_{i})\biggl(1-{1\over{\xi(N,M_{*})}}\biggr)+{NM_{*}
\over{V_{box}\xi(N,M_{*})}}}\biggr]
\end{equation}

Since the mass-energy is conserved,  $NM_{*}=M$, and we obtain

\begin{equation}
\Delta{S_{1}}-\Delta{S_{2}}=M^{2}\biggl(1-{1\over{N}}\biggr)
\end{equation}

since $N > 1$ by hypothesis, this difference is positive.

Using eq.(8) we may express the difference of the total entropies
between the single large black hole and $N$ black holes as

\begin{equation}
\Delta S_{total}^{1} - \Delta S_{total}^{N} = M^{2} \biggl(1-{1\over{N}}\biggr) +
\Delta S_{pert}^{1} - \Delta S_{pert}^{N}
\end{equation}

Therefore, if the entropy associated to the perturbations (not modelled here) is
$\ll$ than $M^{2} (1 -1/N)$; or if their difference happens to be small than the
latter quantity; the system would prefer to form a single large black hole instead
of $N$ small ones. One way to reverse this would be to require a much larger
{\it variation} in the entropy required to form the $N$ black holes than the
variation of the entropy involved in the formation of the single large one.
Clearly a through evaluation of the unmodelled perturbations is needed to
address the outcome on each physically different case.

\bigskip
\section{Black hole masses allowed in the early universe: standard
(non-inflationary) cosmologies}

\bigskip

The considerations of the former sections were useful to understand the basic
features of PBHs formation due to its
simplicity. This analysis can be extended to the FRW model
without drastic modifications. The only difference arises from
the cosmological expansion described by the
proper FRW dynamics, and the use of a fiducial form (yet to be found
unequivocally) of the Holographic Bound. We stress again that there is no
rigurous proof of the validity of the HB for the universe as a whole as yet,
although we shall use the conservative form of the bound
$S_{BH} \leq A/4$ and assume the particle horizon $R_{ph}$ to set the
area (and hence the entropy) in which PBHs may form. Thus we shall write
$S_{ph}$ for the entropy within the particle horizon.

First we address the important case of subdominant PBHs in the
radiation era, arising from the collapse of primordial
perturbations ($\varrho_{rad} \gg \varrho_{pbh}$). In this case,
the PBH formation does not affect significantly the radiation
temperature when they formed (section 3).

The Friedmann equations (without cosmological constant)
and entropy density are given by

\begin{equation}
H^{2}={8\pi{G}\over{3}}\varrho_{total}-{K\over{R^{2}}}
\end{equation}

\begin{equation}
{\ddot{a}\over{a}}=-{4\pi{G}\over{3}}{(\varrho+3P)}
\end{equation}

By considering a flat universe ($\Omega_{0} = 1$), we shall keep the relations
between time and redshift as simple as possible. The extension to
$\Omega {\not =} \, 1$ models and/or alternative functional forms
of the Holographic Bound is straightforward to perform.

For this model the entropy inside the horizon is $S_{ph}(t_{0})\sim{8\times{10}^{121}}$,
the particle horizon is given by the integral $R_{ph}(t)=a(t)lim_{\epsilon\rightarrow{0}}
\int_{\epsilon}^{t}{dt\over{a(t)}}\sim{ct\over{(1-n)}}$,
for $a(t)\propto{t}^{n}$. The entropy evolves according to  $S_{ph}(t)=S_{ph}(t_{0}){(t/t_{0})}^{2}$,
where $t_{0}$ is the present time. This dependence of $S_{ph}\propto{t}^{2}$ is valid in
both the radiation-dominated era or in the matter-dominated era.

The Holographic Bound hypothesis implies that

\begin{equation}
S_{bh}(M)+S_{rad}<A_{ph}(t)/4\sim{8\times{10}^{121}}{(t/t_{0})}^{2} \, .
\end{equation}

The black hole contribution may be written as $S_{bh} =
2.5\times{10}^{40}N{\mu}^{2}$ for a delta-type IMF in which all
black holes have the same mass and $\mu=(M/10^{15} g)$. If we
divide eq.(25) above by the particle horizon volume, the maximal
numerical density in black holes (${cm}^{-3}$) is

\begin{equation}
n_{pbh}(t)<{3\times{10}^{-125}\over{{\pi}}{\mu}^{2}{(t/t_{0})}^{3}}\biggl[8\times{10}^{121}{(t/t_{0})}^{2}-S_{rad}\biggr]
\end{equation}

which may or may not render  relevant limits depending on the difference in brackets.

Since we are considering $\Omega_{0}=1$, then $a(t)=1/(1+z)$ and
$a(t)\propto{t}^{1/2}$, therefore $t(z)=t_{0}{(1+z)}^{-2}$ and $H(z)=H_{0}{(1+z)}^{2}$.
Therefore $n_{pbh}(z)$ becomes

\begin{equation}
n_{pbh}(z) < {3\times 10^{-125}\over {{\pi}{\mu}^{2}}} (1+z)^{6}
\biggl[8\times{10}^{121}{(1+z)}^{-4}-S_{rad}\biggr] \, .
\end{equation}

Multiplying $n_{pbh}(z)$ by the PBH mass and dividing by the critical density
$\varrho_{c}(z)=\varrho_{c}(0){[H(z)/H_{0}]}^{2}$, we finally obtain

\begin{equation}
\Omega_{pbh}(t) < {8 \times{10}^{40}\over{\mu}} \biggl[ 1 \, - \,
{\biggl( {S_{rad}\over{10^{88}}}\biggr)} {\biggl( {(1 +
z)^{4}\over{8 \times 10^{33}}}\biggr)}\biggr]
\end{equation}

At a certain redshift $z_{*}$, $F(z)=0$ and PBHs can not be
formed. This value is $z_{*} \sim {3\times{10}^{8}}$ An inspection
to eq.(28) shows that small, but non-zero values of $\Omega_{pbh}$
can be obtained if PBHs formed at redshifts $z$ very near $z_{*}$,
because the prefactor is large for all realistic values of $\mu$.
Therefore, in practice it is $z_{*}$ the redshift that sets a
relevant limit. For $z > z_{*}$ the HB method is unable to provide
any useful limit to $\Omega_{pbh}$ unless extreme fine-tuning is
invoked.

To close this analysis we stress that we have used the solution
$a(t)\propto{t}^{1/2}$, valid when $\varrho_{rad}(t) \gg
\varrho_{pbh}(t)$ at any given $t$ in the radiation-dominated era.
However, this does not automatically guarantee that subdominant
PBHs do not exceed  the {\it entropy} budget. Therefore, we must
find the conditions for $\varrho_{rad}(t) \gg \varrho_{pbh}(t)$
and $S_{pbh} \sim {S_{rad}}$ to be valid simultaneously. Using
that $\varrho_{rad}(t)\sim{8.4 \times{10}^{5}{(t/s)}^{-2} g
{cm}^{-3}}$ and $\varrho_{pbh}(t)=M_{pbh}n_{pbh}(t)$, the
radiation dominates the expansion when

\begin{equation}
t(\mu) > 1 \times{10}^{-24}{\mu}N_{pbh} \, s
\end{equation}

where $N_{pbh}$ is the total number of PBHs contained within the particle horizon at $t$. However,
the condition $S_{pbh}\sim{S_{rad}}$ holds only if $N_{pbh}{\mu}>
{4\times{10}^{47}/{\mu}}$. Combining both we obtain

\begin{equation}
t(\mu)>{4.7\times{10}^{23}}{\mu}^{-1} \, s \, .
\end{equation}

For consistency we demand this cosmological time to be within the
radiation-dominated era, therefore we must have
$t(\mu)<t_{D}\sim{3\times{10}^{13}s}$, which implies a lower limit
to the mass of the PBHs

\begin{equation}
M_{pbh} > 1.4\times{10}^{26} g \, \sim \, 10^{-7} \, M_{\odot}
\end{equation}

The derived bounds are then actually useful to limit PBHs formed
inside the radiation-dominated era. Our strongest constraint
(always without considering inflation, $\Omega_{0}=1$ and within
the radiation-dominated era) is that above
$z_{*}\sim{3\times{10}^{8}}$ we can not form PBHs, irrespective of
the initial mass, provided they satisfy eq.(31). Before this
value, the Holographic Bound precludes a PBH formation due to the
big entropy of black holes, since the total entropy at disposal is
still small. However, since we have not yet considered inflation
(which is known to stretch the horizon by several orders of
magnitude, among other effects), there may be substantial
modifications to be discussed. The bounds within inflationary
cosmologies will be the subject of next section.

\bigskip

\section{Bounds on PBH masses within inflationary cosmologies}

\bigskip

As is well-known, inflation is a brief stage in the evolution of the universe in which the dynamics
dictates a huge increase in the horizon size; and thus in the entropy within it. We begin by discussing
a constraint on the radius of the particle horizon derived from the same considerations as before (i.e.
comparing entropies before and after the formation of $N$ PBHs) subject to the Holographic Bound. We assume
that before the PBHs formation
$S_{rad}(T(t))<S_{ph}(t)\sim{8\times{10}^{121}}{(t/t_{0})}^{2}$ holds. At $t > t_{f}$, we have
$S_{rad}(T^{\prime})+N_{pbh}S_{bh}(\mu)<S_{ph}(t)$, where $T^{\prime}$ is the lower
temperature due to the combined
effects of cosmological expansion and the PBH formation (section 3).

It is generally difficult to evaluate the two effects together,
but in the approximation in which $\varrho_{pbh} \ll
\varrho_{rad}$ at $t=t_{f}$, we can approximate $T^{\prime}$ by
$T^{\prime}\sim{T{[1-{\varrho_{pbh}\over{\varrho_{rad}}}]}^{1/4}}
\sim{T(1+{\varrho_{pbh}\over{4\varrho_{rad}}})}$. The relation
$N_{pbh}S_{bh}(\mu)+S_{rad}(T(t)) < S_{ph}(t)$ can be inverted to
yield the maximum temperature after the $N_{pbh}$ PBH formation

\begin{equation}
T_{max}(\mu,t)\sim{{\alpha}{(t/t_{0})}^{2/3}\over{(R/cm)}}
\biggl[1-{\varrho_{pbh}\over{8\varrho_{rad}}}\biggr]
{\biggl[1-{{\beta}N_{pbh}{\mu}^{2}\over{{(t/t_{0})}^{2}}}\biggr]}^{1/3}K
\end{equation}

with $\alpha\sim{4.3\times{10}^{40}}$ and $\beta\sim{3\times{10}^{-82}}$.
Since this quantity must be positive, we must have

\begin{equation}
t(N_{pbh},\mu)>1.7\times{10}^{-24}{\mu}\sqrt{N_{pbh}} \, s  \, ,
\end{equation}

to be compared with eq.(30). It may be useful to evaluate $S_{ph}$
in terms of the product $E_{ph}(t)$, the total energy contained
inside the particle horizon times its effective radius
$R_{ph}(t)$. By doing this, we find the alternative formula for
the maximal temperature (we omit hereafter the small correction
term ($1-{\varrho_{pbh}\over{8\varrho_{rad}}}$))

\begin{equation}
T_{max}(N_{pbh},\mu)\sim{{\gamma}\over{(R/cm)}}
{\biggl[{{\lambda}{\varrho_{total}(t){(R/cm)}^{4}}-2.5N_{pbh}{\mu}^{2}}\biggr]}
^{1/3} K \, ,
\end{equation}

with $\gamma\sim{2\times{10}^{13}}$ and $\lambda\sim{8\times{10}^{-2}}$.
Positivity of this temperature implies that

\begin{equation}
\varrho_{total}(t)> {31 N_{pbh}{\mu}^{2}\over{{(R/cm)}^{4}}}g{cm}^{-3}
\end{equation}

On the other hand, we know that the physical temperature in the radiation-dominated
era behaves as

\begin{equation}
T_{rad} \sim{1.2\times{10}^{29}K\over{(R/cm)}}
\end{equation}

PBH formation requires $T_{rad}(t_{f})<T_{max}(t_{f})$, since
that $S_{rad}(t_{f}))<S_{ph}(t_{f})$, we obtain
an additional constraint on the total density for the PBH
nucleation to occur, namely

\begin{equation}
\varrho_{total}(t_{f})\sim{\varrho_{rad}(t_{f})}<
\biggl[{2\times{10}^{48}+31N_{pbh}{\mu}^{2}\over{{(R/cm)}^{4}}} \,
g cm^{-3} \biggr]
\end{equation}

The relation eq.(37) determines a possible range for PBH formation
if we know $\varrho_{rad}(t)$ independently. In the
radiation-dominated era

\begin{equation}
\varrho_{rad}(t)\sim{8\times{10}^{5}}{(t/s)}^{-2} g{cm}^{-3}
\end{equation}

and we know that $R(t)$ behaves as
$R(t)=R_{i}a(t)\propto{t}^{1/2}$, where $R_{i}$ is the horizon
radius at $t_{i}$, well before the PBH formation time (we assume
that a possible inflationary phase has ended much earlier to avoid
complications). $R_{i}$ corresponds to the beginning of the radiation-dominated era
and its value was determined by the end of the inflationary era that
connects smoothly with the latter.

Solving the eqs.(37) and (38) we find that the minimum initial
radius for PBH formation to happen is

\begin{equation}
R_{i} > {8\times {10}^{46}\over{N_{pbh}^{3/4}{\mu}^{3/2}}}\,cm
\end{equation}

Clearly, this imposes some conditions to the number of $e$-folds
of the inflation, in order to produce enough entropy given by
$R_{i}\sim{L_{pl} \, \exp[H \Delta{t}]}$. At a given formation
time of the PBHs, the number of e-folds enough to solve the usual
cosmological problems and to form the PBHs simultaneously must be
at least (using eq.39)

\begin{equation}
N_{e-folds} > 190 - \ln (N_{pbh}^{3/4}{\mu}^{3/2})
\end{equation}

The requirement of an inflationary era is necessary in order to get enough
entropy in the radiation and thus
some room to form Primordial Black Holes. Specific theories with different e-folds will
render different abundances in PBHs at the end of the reheating.

\subsection{PBHs formed inside the inflationary era}

Let us evaluate the surviving PBHs abundances at the end of an
inflationary era. We assume that inflation ends with a reheating
phase where the inflaton decay strongly heats the cosmic
environment. The beginning of the inflation happens when a
classical patch of the space-time is filled with some scalar field
satisfying $V(\varphi)<M_{pl}^{4}$,
$\partial^{\mu}\varphi\partial_{\mu}\varphi<V(\varphi)$. It is
assumed that the initial region is larger than the Planck length
($R_{0}\sim{L_{pl}}$) to avoid dealing with a still unknown theory
of quantum gravity. In a DeSitter phase the scale factor evolves
according to $a(t)=a_{i} \exp(Ht)$, with $H\sim{cte}$. The
particle horizon in this model will be given by
$R_{ph}(t_{f})\sim{L_{pl} \exp(N_{e})}$, where $t_{f}$ corresponds
to the end of the inflation. The entropy at that moment
$S_{ph}(t_{f})$ will be

\begin{equation}
{A_{ph}(t_{f})\over{4L_{pl}^{2}}}\sim{\pi{\exp(2N_{e})}}
\end{equation}

As far as we know, there is no upper bound to $N_{e}$, but it must
be $\geq {67}$ to get rid of the well-known problems such as
homogeneity, horizon, etc. (see [12]). The difference of the
Holographic Bound entropy and the radiation entropy (produced by
the decay of the inflaton field and of order $\sim{10}^{88}$) will
be $S_{bh}(\mu)=CN_{pbh}{\mu}^{2}={\pi}exp(2N_{e})-S_{rad}$, where
$C\sim{2\times{10}^{40}}$. Following the same reasoning of the
former sections, we obtain for the PBH density

\begin{equation}
\varrho_{pbh}(t_{f})\leq{3\times{10}^{15}F(N_{e})\over{4{\pi}}{\mu}{L_{pl}}^{3}C}
\end{equation}

with $F(N_{e})={\pi} \exp(-N_{e})-S_{rad} \exp(-3N_{e})$.

Today $\varrho_{c}(t_{0})\sim{10}^{-29}g{cm}^{-3}$, but at $t =
t_{f}$ we only know that
$\varrho_{c}(t_{f})<{M_{pl}}^{4}\sim{10}^{93}g{cm}^{-3}$ (see
[14]). A parameter $\eta < 1$ may be introduced to account for
these unknown details as
$\varrho_{c}(t_{f})={\eta}{10}^{93}g{cm}^{-3}$. Using this
definition, an upper limit to $\Omega_{pbh}$ results, namely

\begin{equation}
\Omega_{pbh}(\mu,N_{e})<6\times{10}^{-21}{F(N_{e})\over{\eta\mu}}
\end{equation}

It is clear that $F(N_{e})=0$ for
$N_{e}^{*}={1\over{2}}ln{({S_{rad}\over{\pi}})}\sim{100.7}$. Then,
for $N_{e}$ below $\sim{101}$ no PBHs can be formed at $t_{f}$.
The function $F(N_{e})$ has a maximum at
$N_{e}^{max}={1\over{2}}ln{({3S_{rad}\over{\pi}})} \sim{101.3}$,
and the maximal value for $F(N_{e})$ is $\sim{2\times{10}^{-44}}$.
Then, the maximal abundance of $\Omega_{pbh}$ (at $t_{f}$) will be

\begin{equation}
\Omega_{pbh}(t_{f})< {1.2\times{10}^{-64}\over{\eta\mu}}
\end{equation}

The volume of the horizon after $t_{f}$ will be even larger than
$\sim{{L_{pl}}^{3} \exp(3N_{e})}$ and $\Omega_{pbh}$ drops even
more in these conditions. There is no hope of producing a sizeable
$\Omega_{pbh}$ of PBHs formed inside the inflationary era. The
reasons for this huge PBH dilution are similar to that found in
Guth's analysis of the magnetic monopoles abundances [13]. The
huge injection of entropy (and its associated expansion) dilutes
the magnetic monopole abundances to negligible values
$\Omega_{mm}\sim{10}^{-81}$ or less. For PBHs, the effect is
analogous, but as these objects are extremely entropic
($C\sim{10}^{40}$), we obtain the bounds of eq.(44) which are
equally strong for $\sim \, 1 \, M_{\odot}$ black holes.
Nevertheless, the dilution is always very strong and states that
PBHs do not contribute to $\Omega_{total}$ if formed within the
inflation, as expected.

\subsection{PBHs formed after an inflationary era}

The inflation of $R_{ph}$ proved to be lethal for PBHs formed
inside the inflation, but the former certainly created a favorable
setting after $t_{f}$ for any process that may induce PBH
formation. We can evaluate the maximal PBH abundances at $t>t_{f}$
as follows. We will simplify the model by assuming that the
inflationary phase lasts $\Delta t \leq {t_{f}}$, and from
$t_{f}=\Gamma\times{10}^{-35} s$ on (with $1<\Gamma<{10}^{3}$
being a model-dependent parameter including the unknown details of
the epoch) the universe entered the radiation-dominated era Then
$a(t)\propto \exp(t)$ for $t < t_{f}$, and after this phase, the
decay of the inflaton field reheats the background as the universe
enters the radiation-dominated era in which
$a(t)\propto{t}^{1/2}$. The numerical details of this transition
are not important for our estimates, as long as it occurs
immediately after $t_{f}$. The continuity of the horizon particle
allows us write $R_{ph}(t)={ct_{f}\over{(1-n)}}(t/t_{f})$. This
value is adopted as the initial $R_{i}$ of the radiation-dominated
era $R_{ph}(t_{f}) \equiv R_{i}={ct_{f}\over{(1-n)}}=L_{pl}
\exp(N_{e})$.

Therefore we have $R_{ph}(t)\sim{L_{pl} \exp(N_{e})}(t/t_{f})$ for
$t>t_{f}$, and the entropy evolves as

\begin{equation}
S_{ph}(t)={\pi} \exp(2N_{e}){(t/t_{f})}^{2}
\end{equation}

Following the same reasoning as above, we write now for $\Omega_{pbh}$

\begin{equation}
\Omega_{pbh}(\mu,N_{e},t)< 9.4\times
10^{-4}{{\Gamma}^{2}\over{\mu}} {(t/t_{f})} \times G(N_{e},t) \,
\times \exp{(-N_{e})}
\end{equation}

where

\begin{equation}
G(N_{e},t)=1-{S_{rad}\over{\pi{(t/t_{f})}}^{2}} \exp(-2N_{e})
\end{equation}

valid only for $1< (t/t_{f}) < (t_{D}/t_{f}) \sim{(3/{\Gamma})}{10}^{48}$, that is,
from the end of inflation till the beginning of the matter-dominated era.

We can proceed to evaluate the maximal abundances in PBHs formed
at a time $t_{form}$. Most of the models that describe PBH
formation from primordial fluctuations (see [15]), estimate that
the PBH mass at its formation, $t_{form}$, is a fraction $\beta
\leq 1$ of the particle horizon. Then, it is  reasonable to put

\begin{equation}
\mu(t=t_{form})\sim{\beta}(M_{hor}(t_{form})/M_{haw})
\sim{9\beta\times{10}^{22}(t_{form}/s)}
\end{equation}

Furthermore, as stated above the initial conditions for inflation
are $V(\varphi)>>\partial^{\nu}\varphi\partial_{\nu}\varphi$ and
$V(\varphi)<M_{pl}^{4}$. The last one says that the spacetime is
classical. We do not know exactly the initial energy density
stored by the inflaton field, but it is generally assumed that
$V(\varphi)={\eta}M_{pl}^{4}$, with $\eta<1$. Using the Friedmann
equation $H_{0}^{2}\propto{V(\varphi)}$, $H_{0}t_{f}=N_{e}$  and
recalling the definition of $t_{f}$ we identify
$t_{f}=\Gamma\times{10}^{-35}s
\sim{2.2\times{10}^{-44}{N_{e}\over{\eta^{1/2}}}} s$. Substituting
$\mu$ and $\Gamma$  into eq.(46) (evaluated at $t=t_{form}$)
yields

\begin{equation}
\Omega_{pbh}(N_{e},t=t_{form})< {2.1 \times
N_{e}\over{{\beta}{\eta^{1/2}}}} \, \times \, G(N_{e},t_{form}) \,
\times 10^{-{43\over{100}} N_{e}}
\end{equation}

The PBH abundance is zero when the function $G(N_{e},t_{form})$ vanishes,
this happens at

\begin{equation}
t_{\ast} \sim {0.9 N_{e}{\eta}^{-1/2} \exp(-N_{e})} \, s
\end{equation}

All those inflationary models with a large number of e-folds
$N_{e}$, will not form PBHs at times earlier than $t_{\ast}$.
After $t_{\ast}(N_{e},\eta)$, the difference between the total
entropy (given by the GSL) and the Holographic Bound will allow
the PBH formation by collapse of large fluctuations. Other PBH
formation mechanism(s) allowing masses much below $M_{hor}$ will
suffer milder restrictions, yet to be studied.

\section{Conclusions}

After analyzing the formation of PBHs inside boxes and possible
extensions to realistic cosmological models, we conclude that if
the Holographic Bound holds, we can use it to obtain constraints
on the black hole formation and other features (see the reviews in
[16] and [17] for complete discussions of the Holographic
Principle). The usefulness of this method is rooted on the huge
value of the black hole entropy, which poses limitations to
uncontrolled formation of PBHs. Furthermore, since the early
universe was very small (compared to the present universe), the
total entropy within it was much smaller than today and useful
limits can be extracted from some upper bound to the total
entropy. This method can be applied to constrain the epochs and
masses of PBH nucleation and new requirements on the early
conditions obtained. It is not a feature of a particular form of
the Holographic Principle, but rather a general feature that can
be implemented for any given form of the latter, and which may
prove restrictive for their abundances in several cases. We
obtained useful bounds for standard (non-inflationary) and
inflationary cosmological models. For the former a threshold
redshift $z_{*}$ quenching the formation of PBHs has been found;
while the limits are strong for the latter in all cases. The
minimal set of assumptions made for inflationary cosmologies
suggests that this low abundance of PBHs is a quite generic
feature of the models.

All these cosmological applications made use of the expressions
$S_{BH}$ and $S_{BV}$,  without considering, for example, the
influence of the equation of state. We may rederive all these
results again, considering any generalization of them, such as the
proposal made by Youm [18]. The {\it growth} of PBHs has been also
dismissed in all this work. However, it is known that the black
holes formed in the very early universe can grow at expenses of
the radiation around them (see [19] ) and [20]). Note that as PBHs
grow with time, the total entropy of the universe grows
proportional to their masses. A simplified analysis of the problem
of PBHs growth has been presented recently by the authors [21]
using model-independent arguments. In all the latter analysis it
has been found that these PBHs (almost) did not grow at all, and
therefore, the maximum of the entropy was achieved practically at
the instant of their formation. In spite that we expect small
modifications to the present results when the growth of PBHs is
included, a detailed analysis is needed to confirm this
expectation.

\section{Acknowledgements}

Both authors wish to thank the S\~ao Paulo State Agency FAPESP for
financial support through grants and fellowships. J.E.H. has been
partially supported by CNPq (Brazil). We would like to thank the
scientific advice from E. Abdalla on several aspects of this work.

\section{References}

1) S.W.Hawking, {\it Comm.Math.Phys.}{\bf 43}, 199 (1975).

2) B.Carr, {\it Astron. and Astrophys. Transactions} {\bf 5}, 43 (1994).

3) A. R. Liddle and A. M. Green, {\it Phys. Reports}{\bf 307}, 125 (1998);
A.M. Green and A.R. Liddle {\it Phys. Rev. D}{\bf 60}, 063509
(1999).

4) P.S.Custodio and J.E.Horvath, {\it Phys. Rev. D}{\bf 65},
024023 (2002).

5) B.Carr, {\it Phys. Reports} {\bf 307}, 141 (1998).

6) J.Bekenstein, gr-qc/9409015 (1994).

7) T.Jacobson, gr-qc/9801015 (1998).

8) see for example S. Mukohyama, gr-qc/9912103 (1999) and references therein.

9) E.Verlinde, hep-th/0008140 (2000).

10) M.G. Ivanov and I.V. Volovich, gr-qc/9908047 (1999).

11) J.Bekenstein, {\it Phys. Rev. D}{\bf 9}, 3292 (1974).

12) E.W. Kolb and M.S. Turner, {\it The Early Universe}, (Addison-Wesley, Reading MA 1990)

13) A. H. Guth, in {\it Magnetic Monopoles}
(Proceedings of the NATO Advanced Study Institute), eds. R.A.
Carrigan and W.P. Trower (Plenum, NY 1983)

14) A. Linde, {\it Particle Physics and Inflationary
Cosmology} (Harwood Academic Press, NY 1990)

15) B.Carr, {\it Astrophys.J}{\bf 201}, 1 (1975).

16) R. Bousso, hep-th/0203101 (2002).

17) D.Bigatti and L.Susskind, {\it TASI lectures on the Holographic
Principle}, hep-th/0002044 (2000).

18) D.Youm, hep-th/0201268 (2002).

19) Ya. B. Zel'dovich and I.D. Novikov, {\it Relativistic Astrophysics}
(Univ. Chicago Press, Chicago 1983).

20) B.Carr and S.W.Hawking, {\it MNRAS} {\bf 168}, 399 (1974).

21) P.S. Cust\'odio and J.E. Horvath {\it Phys. Rev.D}{\bf 58}, 023504
(1998) ; {\it ibid}, {\it Gen. Rel. Grav.}{\bf 34}, 1895 (2002).

\vfill\eject

\noindent Figure captions.

\bigskip
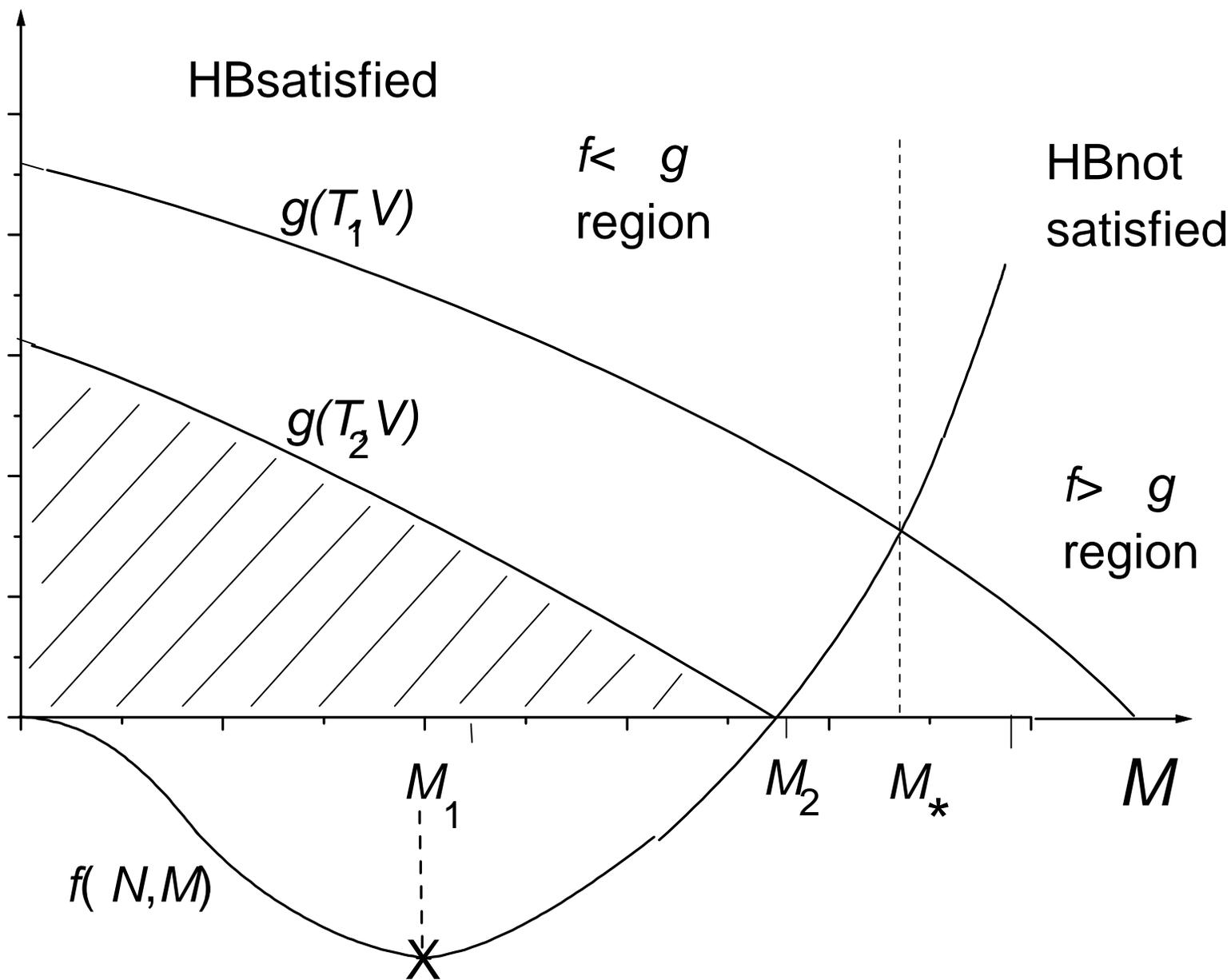
\noindent Fig. 1. Graphical solution of the formation of PBHs at
the expense of background radiation within a closed box. The
curves $g(T_{1}, V_{box})$ and $g(T_{2}, V_{box})$ refer to two
different cases. In the first, $T_{1}>>T_{2}$ and the formation of
the black holes does not exhaust the radiation within the box
(upper solid curve). In the second case (lower solid curve), the
formation of black holes exhausts the ambient radiation
completely. As discussed in the text, the Holographic Bound is
more restrictive than the total content of energy to set the
maximal mass for individual black holes; unless the conditions in
the box are such that $M_{max} < M_{2}$ (with $M_{2}$ the zero of
the function $f (N,M)$), since in the latter situation the energy
constraint is stronger than the Holographic Bound to evaluate the
PBH formation.

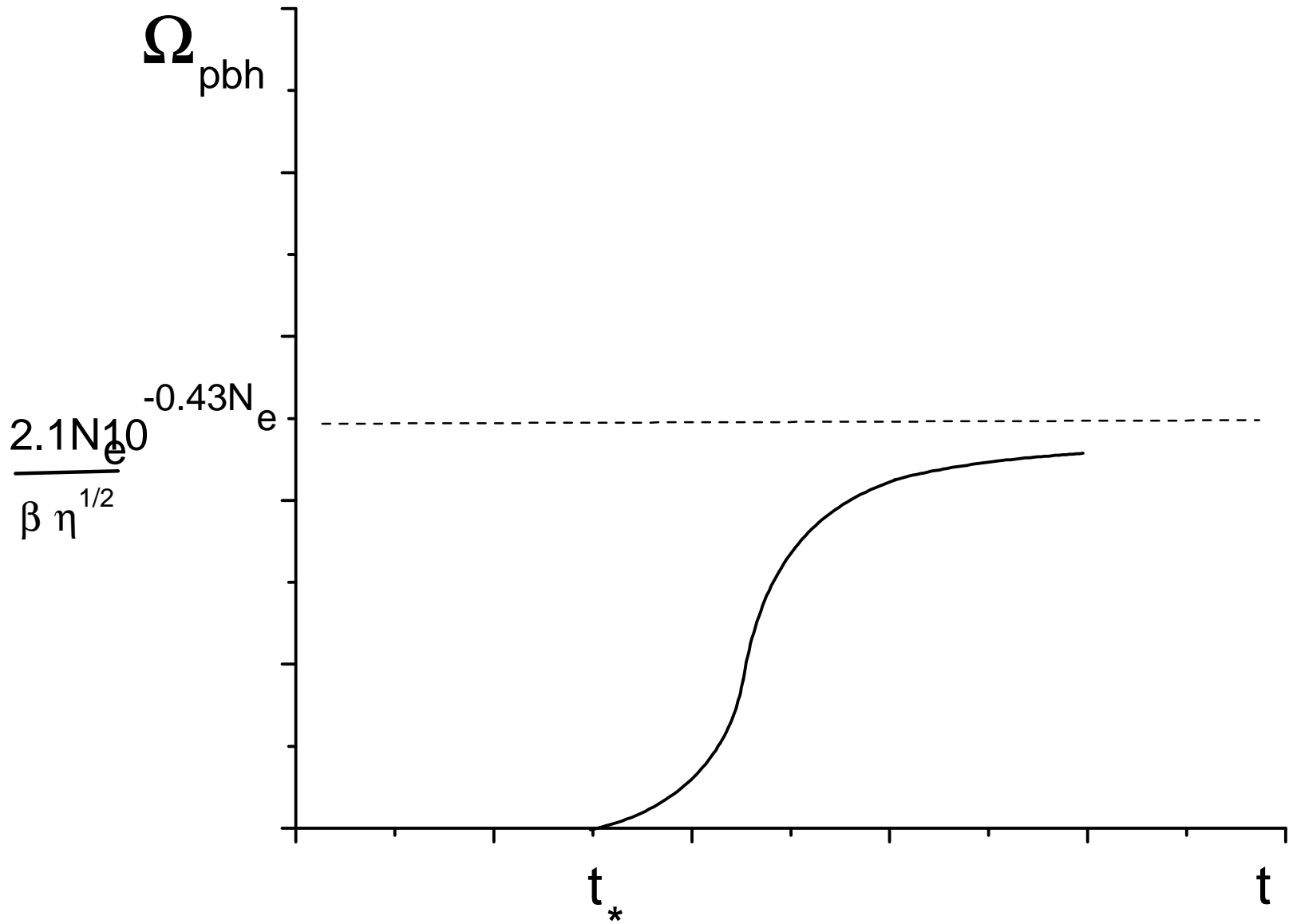
Fig. 2. Primordial black hole abundance limits within inflationary
models. As discussed in the text,$\Omega_{pbh} \equiv 0$ before
$t_{*}$. Note that even for PBHs formed at asymptotic times (that
is, when $G \rightarrow 1$) the allowed contribution to
$\Omega_{pbh}$ is very small because of the suppression by $N_e$.
Physically the suppression reflects the fact that the volume grows much
more than the entropy inside the particle horizon.

\end{document}